# Targeted-Subharmonic-Eliminating Pulse Density Modulation for Wireless Power Transfer System

Songyan Li, *Graduate Student Member, IEEE,* Hongchang Li, *Senior Member, IEEE*

*Abstract*—This letter proposes a targeted-subharmonic-eliminating pulse density modulation (TSE-PDM) method for SS-compensated WPT systems. By designing a noise transfer function with notch characteristics, the subharmonic components which excite current abnormal oscillations were eliminated. Simulation and experimental results demonstrate the effectiveness of the TSE-PDM in suppressing current abnormal oscillations. The proposed method is easy to implement in either primary or secondary side of the WPT system and exhibits a certain tolerance to deviations in NTF design, representing the most straightforward method for abnormal oscillation suppression in PDM controlled WPT systems.

*Index Terms*—Wireless Power Transfer (WPT), pulse density modulation (PDM), subharmonic, current oscillation, zero-voltage switching (ZVS).

## I. INTRODUCTION

Pulse density modulation (PDM) is a promising approach for high-efficiency power conversion, which has become a focus of research in wireless power transfer (WPT) systems. Unlike Phase Shift Modulation (PSM) suffering from hard-switching, PDM enables wide-range gain adjustment in both primary and secondary side, while consistently maintaining soft-switching operation with low reactive power and reduced the average switching frequency.

Delta-Sigma Pulse Density Modulation (ΔΣ-PDM) enables continuous modulation with theoretically arbitrary pulse density resolution, and is ideal for efficiency-optimized closed-loop control systems. However, experimental results reveal that subharmonic component in the modulated wave can unexpectedly excite abnormal oscillations in resonant current under specific conditions, characterized by periodic significant fluctuations in the envelope of resonant currents. Thus, various solutions have been proposed to solve this problem. In [3], an active damping control by transiently introducing PSM is proposed to suppress the current oscillations, but it causes hard-switching in some cycles. In [4], by adding a sending current limiter as a condition for skipping pulses, the conditional PDM suppresses current oscillations while maintaining soft-switching operation, but the limiting current thresholds under other coupling coefficients have not been systematically discussed. As presented in [5], the rectifier and inverter are synchronously modulated with precisely controlled pulse density and phase alignment, leveraging transient response superposition to eliminate oscillations, but precise current envelope detection is required to identify the pulse-skipping instances.

This letter analyzes the essential reasons for the abnormal oscillation in series-series (SS) compensated WPT system caused by ΔΣ-PDM, and proposes the targeted-subharmonic-eliminating PDM (TSE-PDM). The proposed modulator employs a deliberately engineered Noise Transfer Function (NTF) with notch characteristics to eliminate subharmonic components in the modulated wave that excites abnormal oscillations.

## II. ANALYSIS FOR CURRENT OSCILLATION

The typical circuit diagram of the SS-compensated WPT system is shown in Fig. 1, where $V_g$ ($V_o$), $U_1$ ($U_2$), $I_1$ ($I_2$), $L_1$ ($L_2$), $C_1$ ($C_2$) and $R_1$ ($R_2$) are the input (output) voltage, primary (secondary) voltage modulated wave, current, inductance, capacitance, and equivalent series resistance, respectively. $M$ is the mutual inductance while $k$ is the coupling coefficient.

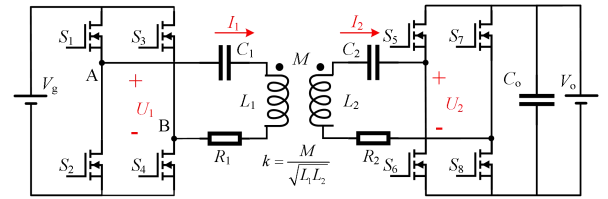

**Fig. 1.** Circuit diagram of the SS-compensated WPT system.

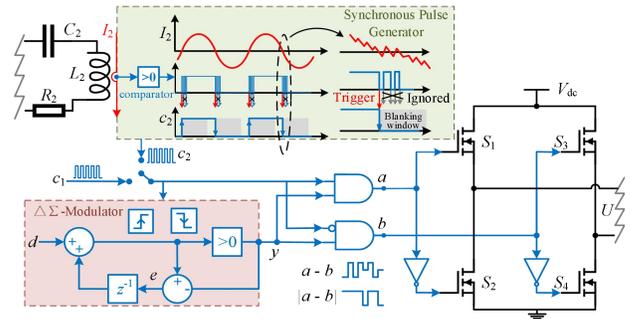

**Fig. 2.** Structure of the conventional ΔΣ-PDM converter.

The structure of a conventional ΔΣ-PDM converter implemented on the primary/secondary-side of the WPT system is depicted in Fig. 2. In each half-switching-cycle, the quantization error $e$ accumulates with the input pulse density $d$, and the comparator rounds the accumulated results. The output of quantizer $y$ is AND-ed with the pulse $c_{1/2}$ and the inverted pulse to generate the driving signals for the leading-leg and lagging-leg switches. $c_1$ is the primary input pulse, while $c_2$ is the secondary synchronous pulse for rectification. In synchronous pulse generator, the sampled current signal is compared with zero, and the edges of the square wave generated by comparator trigger the transition of synchronous pulse $c_2$. To avoid false triggering events, a blanking window is generated after the transition of $c_2$, and the series of edges is ignored while happening during the given blanking period.

The spectrum of modulated wave $a - b$ contains abundant non-fundamental frequency components that are non-integer multiples of the switching frequency $\omega_s$, namely subharmonic. Due to the frequency-selective characteristics of the SS-compensated network, the subharmonic components are effectively suppressed, and the current typically exhibits minor fluctuations. However, under specific pulse density conditions, the current demonstrates abnormal oscillations exceeding 50% in amplitude as shown in Fig. 3. The abnormal oscillations increase conduction losses, disrupt soft-switching operation, and may potentially lead to damage in capacitors and switching devices.

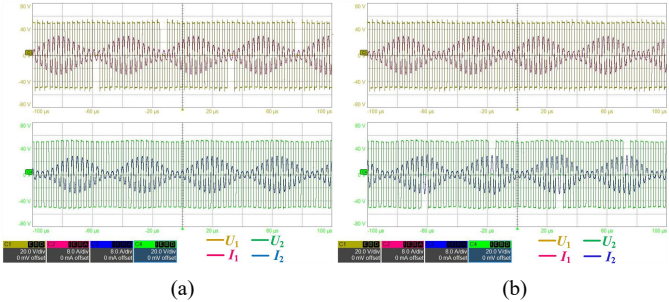

Fig. 3. Experiment results with abnormal oscillations of the WPT system ($k$ = 0.152) under conventional ΔΣ-PDM control: (a) operation at $d_1$ = 0.963 while $d_2$ = 1, and (b) operation at $d_2$ = 0.963 while $d_1$ = 1.

The spectrum of modulated wave exhibits symmetry about the fundamental frequency $\omega_s$. A pair of subharmonics at frequencies $\omega_s \pm \Delta\omega$ can be equivalently represented as an amplitude-modulated fundamental wave, as:

$$\cos[(\omega_s + \Delta\omega)] + \cos[(\omega_s - \Delta\omega)] = 2\cos(\Delta\omega)\cos(\omega_s t) \quad (1)$$

Therefore, the subharmonic excitation can be equivalently converted into fundamental-wave excitation with low-frequency amplitude modulation. The generalized state-space averaging (GSSA) model is employed to investigate the transfer function from voltage amplitudes to resonant current amplitudes, providing the reason of abnormal oscillations caused by specific subharmonics.

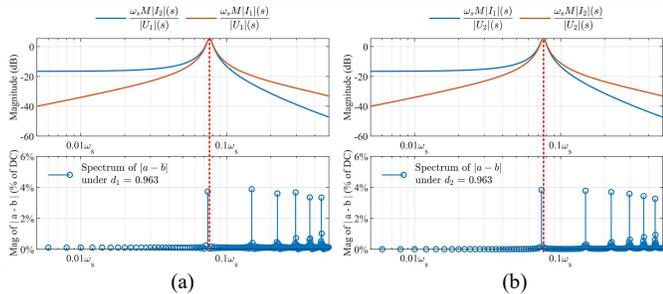

Fig. 3. Comparison between the bode plots of voltages to currents at $k$ = 0.15 and the spectrum of the modulated wave amplitude |a - b| at $d_{1/2}$ = 0.963: (a) the bode plot of $U_1$ to $I_1$ and $I_2$ and the spectrum of |a - b|, and (b) the bode plot of $U_2$ to $I_1$ and $I_2$ and the spectrum of |a - b|.

In Fig. 3, the small signal model of transfer function from the voltage amplitude to the current amplitude at $k$ = 0.15 is obtained through GSSA method, with a resonant peak at:

$$\omega_0 = \frac{1}{2}k\omega_s \quad (2)$$

The spectrum of modulated wave amplitude under $d_{1/2}$ = 0.963 is obtained through Fourier decomposition, as shown in Fig. 3, which contains components located near the resonant peaks of the amplitude transfer function.

Therefore, the abnormal current oscillation cause by ΔΣ-PDM is that the SS-compensated WPT system fails to suppress the frequency components located at $\frac{1}{2}k\omega_s$ in the spectrum of modulated wave amplitude $|a - b|$, or the subharmonic located at $\omega_s \pm \frac{1}{2}k\omega_s$ in the spectrum of modulated wave $a - b$ according to (1).

### III. Proposed TSE-PDM

When the coupling coefficient $k$ is fixed, only particular subharmonics may excite oscillation. A noise shaping method to eliminate the target subharmonic is proposed, which actively suppresses potential oscillation-inducing components within the modulator to prevent abnormal current oscillation fundamentally.

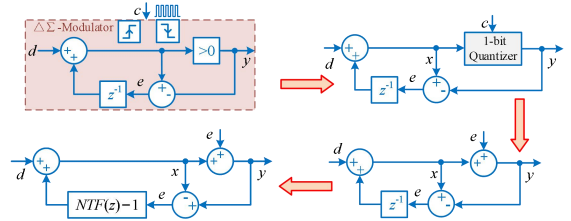

Fig. 4. Derivation of the linearized model for ΔΣ-modulator.

The linearized model of ΔΣ pulse-density modulator is constructed, as depicted in Fig. 4. The quantization error is modeled as an additive noise term $e$, and the final modulator architecture is obtained through equivalent linear network transformation. The quantizer output $y$ equals the modulated wave amplitude $|a - b|$, which satisfies the z-domain equation:

$$Y(z) = D(z) + E(z)NTF(z) \quad (3)$$

where $Y(z)$, $D(z)$, $E(z)$ and $NTF(z)$ represent the z-transform of the quantizer output $y$, input pulse density $d$, quantization error $e$ and NTF, respectively. The NTF of the ΔΣ pulse-density modulator shown in Fig. 4 is a first-order difference block:

$$NTF_1(z) = 1 - z^{-1} \quad (4)$$

To achieve noise shaping with targeted subharmonic suppression, it is necessary to investigate a high-order NTF that differs from (4), which must satisfy the following four requirements:

① The DC gain of the NTF must be adequately low:

$$NTF(1) = 0 \quad (5)$$

② The NTF must be physically realizable, that is the z-domain transfer function NTF(z) - 1 must be zero-initialized:

$$NTF(\infty) = 1 \quad (6)$$

③ The NTF must contain a pair of complex-conjugate zeros at $\omega_e$ that represents the frequency component to be eliminated:

$$z_{1,2} = e^{\pm j\pi\frac{\omega_e}{\omega_s}} \quad (7)$$

④ The NTF must guarantee the stability of the 1-bit ΔΣ-modulator, that is the quantization error $e$ must be within the range of -1 to 0.

A third-order NTF satisfying the mentioned conditions is constructed as:

$$NTF_3(z) = \frac{z-1}{z-0.9} \frac{(z-e^{j\pi\frac{\omega_e}{\omega_s}})(z-e^{-j\pi\frac{\omega_e}{\omega_s}})}{(z-0.9e^{j\pi\frac{\omega_e}{\omega_s}})(z-0.9e^{-j\pi\frac{\omega_e}{\omega_s}})} \quad (8)$$

For the ΔΣ-modulator with 1-bit quantizer, the first-order modulator is unconditionally stable, while the bounds of stability of second-order modulator can be mathematically proven [6]. But the high-order NTF properties that are necessary and sufficient for stable operation is not known. The stability of high-order modulator needs to be confirmed by extensive simulations.

For a completely resonant SS-compensated WPT system with $k = 0.15$, setting $\omega_e = 0.075\omega_s$ can suppress subharmonics that cause abnormal oscillations according to (2). In Fig. 5, the input pulse density $d$ is varied as sinusoidal and ramp waveform. The quantizer output $y$ tracks the input $d$, and the error $e$ always remains within the range of -1 to 0.

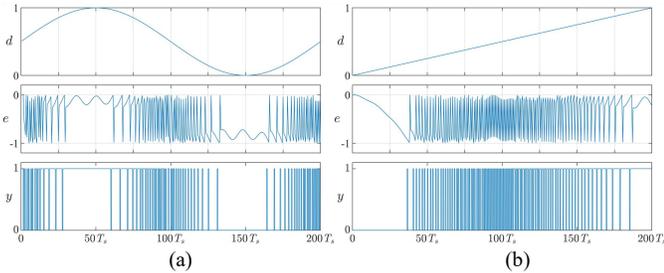

Fig. 5. Dynamic response of the modulator with $NTF_3$ to (a) sinusoidal excitations and (b) ramp excitations, $\omega_e = 0.075\ \omega_s$ (i.e., the case with $k = 0.15$).

Fig. 6 compares the bode plots and pole-zero diagrams of different NTFs, along with the waveform and spectrum of modulated wave generated by modulators applying these NTFs. In the time domain, the different NTFs achieve noise shaping through distinct pulse arrangements while maintaining constant pulse density and unchanged fundamental amplitude, guaranteed by the zero of the NTFs at $z = 1$. With proper pole-zero placement, the $NTF_3$ achieves zero gain at $0.075\omega_s$, enabling the pulse density modulator to eliminate the frequency component near $0.075\omega_s$ in the modulated wave amplitude $|a - b|$. And the paired subharmonic near $0.925\omega_s$ and $1.075\omega_s$ in the spectrum of modulated wave $a - b$ are simultaneously canceled according to (1).

In summary, the proposed TSE-PDM is based on a ΔΣ-modulator with a third-order NTF featuring notch characteristics instead of the conventional first-order difference block. The modulated waveform contains no subharmonics near the resonant peaks of the resonant network's input admittance and output admittance, thus preventing abnormal oscillations.

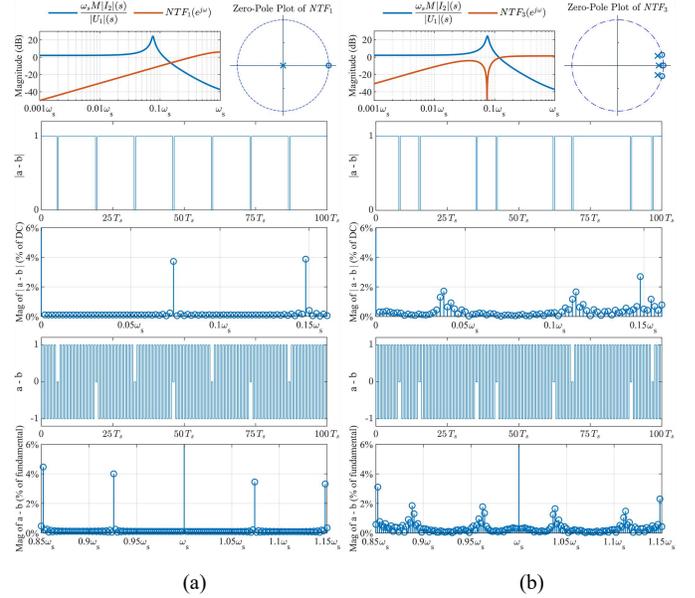

Fig. 6. Comparison of NTFs: bode plots, pole-zero diagrams, modulated waveform and spectrum at $d = 0.963$. (a) The $NTF_1$ (b) The $NTF_3$.

IV. EXPERIMENTAL VERIFICATION

A 220W SS-compensated WPT system is built for verifying the purposed method. The parameters of the prototype are listed in Table I.

TABLE I
WPT SYSTEM PARAMETERS

| Symbol | Quantity | Value |
|---|---|---|
| $L_1, L_2$ | Resonant inductances | 31.7μH, 29.7μH |
| $C_1, C_2$ | Resonant capacitances | 8.87nF, 9.47nF |
| $R_1, R_2$ | ESR | 105mΩ, 102mΩ |
| $V_g, V_o$ | DC side voltage | 50V |
| $k$ | Coupling coeffiicient | 0.152 |
| $f_s$ | Switching frequency | 300kHz |

A. Steady-State Performances Under Ideal Noise Shaping

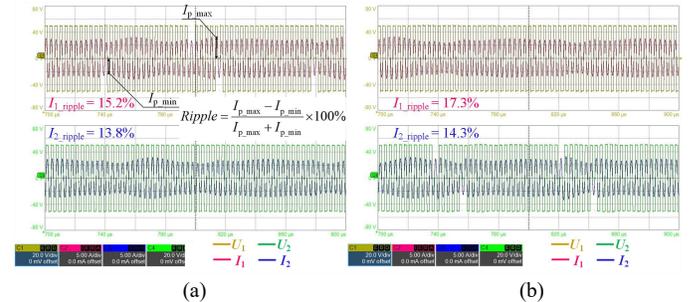

Fig. 7. Experimental waveforms under TSE-PDM at (a) t $d_1 = 0.963$ while $d_2 = 1$ and (b) $d_2 = 0.963$ while $d_1 = 1$.

The steady-state performances of the system under conventional ΔΣ-PDM control and proposed TSE-PDM control were tested at operating points that $d_{1/2}$ changes from 0.203 to 1 while the pulse density of another side is fixed at 1.

The notch frequency of $NTF_3$ is precisely set as the resonant peak frequency $\omega_0$, that is $0.076\omega_s$.

Especially in Fig. 7, the current ripple is reduced from more than 50% to around 17% at $d_{1/2} = 0.963$, compared with Fig.3 which is tested under the same system parameters. And the current ripple across the entire test range of $d_{1/2}$ was suppressed to within 22% as depicted in Fig.8,

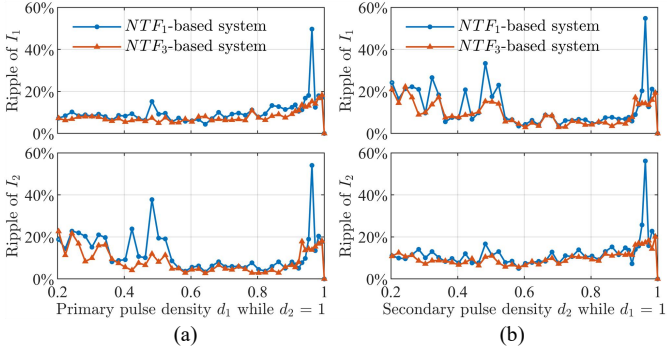

**Fig. 8.** The current ripple of the ΔΣ-PDM system with different NTFs under (a) primary side control and (b) Secondary side control.

### B. Noise Shaping with Certain Deviation

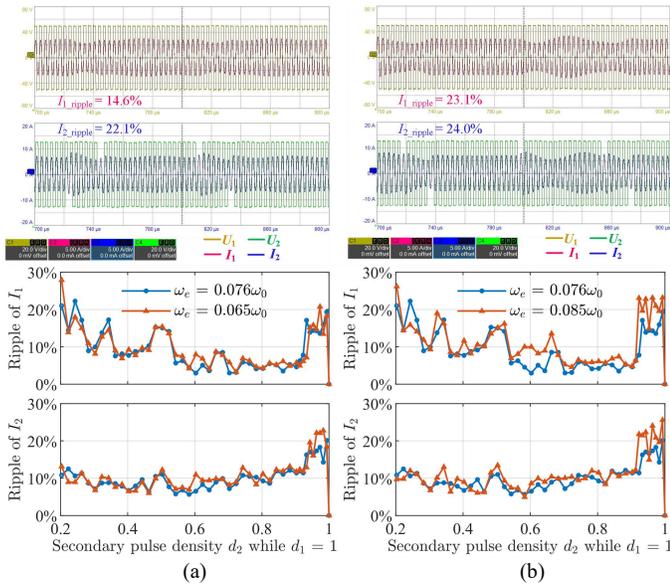

**Fig. 9.** The waveform at $d_2 = 0.963$, and the current ripple under noise shaping with deviation which is compared with the ideal condition. (a) $\omega_e = 0.065\omega_s$. (a) $\omega_e = 0.085\omega_s$.

The proposed TSE-PDM method designs the notch frequency of the NTF based on the coupling coefficient $k$ of coils, but the identification of $k$ inevitably contains errors in practice.

In Fig.9(a), it is assumed that $k$ is inaccurately estimated to be 0.13 with a relatively large error, thus the notch frequency is set to be $0.065\omega_s$ according to (2). Compared to the ideal case, the secondary-side current ripple increases from 17.3% to 22.1% at $d_2 = 0.963$, and the maximum current ripple within the range of $d_2 = 0.9\sim1$ increases from 20.1% to 22.8%. In Fig.9(b), the coupling coefficient is inaccurately estimated to be 0.17, and the notch frequency is set to be $0.085\omega_s$. The current ripple increases to around 23% at $d_2 = 0.963$, and the maximum current ripple within the range of $d_2 = 0.9\sim1$ increases to 25.6%. Since the current ripple remains suppressed within an acceptable range, the proposed TES-PDM method exhibits a certain tolerance to variation or inaccurate estimation of coupling coefficinet.

### C. Sinusoidal Responses

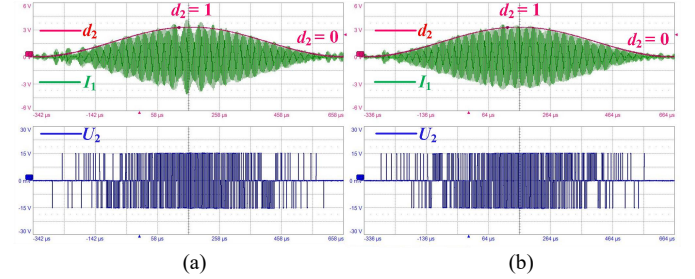

**Fig. 10.** Sinusoidal responses of the ΔΣ-PDM system under different NTFs. (a) $NTF_1$-based system. (b) $NTF_3$-based TSE-PDM system.

Fig. 10 shows the ΔΣ-PDM system's sinusoidal responses when $d_1 = 1$ and $d_2 = 0.5\sin(1000\pi t) + 0.5$ under the conditions of 15V input and output dc voltages. Compared with the conventional $NTF_1$-based system, the current of the TSE-PDM system under ideal noise shaping exhibits reduced oscillation, and the current envelope is basically consistent with $d_2$ (output by DAC), demonstrating that the pulse density of modulated wave $U_2$ effectively tracks variations in $d_2$. These characteristics show that the proposed TSE-PDM strategy exhibits wide modulation range and fast response.

## V. CONCLUSION

This letter proposes a TSE-PDM method based on noise shaping for SS-compensated WPT systems. The subharmonic components that excite abnormal oscillations are eliminated through a third-order NTF featuring notch characteristics embedded in the ΔΣ-modulator. TSE-PDM is a low cost and straightforward method for preventing abnormal oscillations, and it also exhibits certain tolerance to deviations in NTF design and coupling coefficient variation.